\documentclass[preprintnumbers]{revtex4}

\usepackage{graphicx}
\usepackage{amsmath,amssymb}
\begin{document}
\bibliographystyle{apsrev}

\preprint{BNL-HET-01/36}

\title[]{Inclusive Higgs Production at Next-to-Next-to-Leading Order}



\author{Robert V. Harlander}
\email[]{robert.harlander@cern.ch}
\thanks{Work supported by {\it Deutsche Forschungsgemeinschaft}}
\author{William B. Kilgore}
\email[]{kilgore@bnl.gov}
\thanks{Work supported by the United States Department of Energy
 under grant DE-AC02-98CH10886.}
\affiliation{Physics Department\\
             Brookhaven National Laboratory\\
             Upton, New York 11973}

\date{\today}

\begin{abstract}
  We describe the contributions of virtual corrections and soft gluon
  emission to the inclusive Higgs boson production cross section
  $pp\to H + X$ computed at next-to-next-to-leading order in the
  heavy top quark limit.  We also discuss estimates of the leading
  non-soft corrections.
\end{abstract}

\maketitle

\section{Introduction}
The Standard Model is almost thirty-five years old, and its essential
goal, to describe the electro-weak interactions as a spontaneously
broken $SU(2)_L\otimes U(1)_Y$ gauge symmetry has been spectacularly
confirmed.  However, the agent of electroweak symmetry breaking
remains elusive.
The simplest model of symmetry breaking, called the Minimal Standard
Model, uses a single complex doublet of fundamental scalars and is the
benchmark for studies of the symmetry breaking sector of the theory.
Direct search limits from LEP tell us that the Higgs mass is greater
than $\sim114$ GeV.  Fits to precision electroweak data prefer a
mass well below the direct search limit although the $95\%$ confidence
level upper limit is somewhat greater than $200$ GeV.

Higgs boson production at hadron colliders is dominated by the gluon
fusion mechanism.  However, experiments must not only produce Higgs
bosons, they must also detect them.  With a center-of-mass energy of
$2$~TeV, the Fermilab Tevatron is primarily sensitive to a Higgs boson
with mass below the threshold for decay into $W$ boson pairs.  In this
case, the Higgs will decay almost exclusively into $b\bar{b}$ pairs
which will be undetectable on top of an enormous QCD background.
Since the total cross section is too small to permit the use of rare
decay modes, a light Higgs can only be detected through associated
production with a $W$ or $Z$ boson.  Only if the Higgs is sufficiently
massive that the $WW^*$ channel begins to open up, will inclusive
production via the gluon fusion mechanism be useful in the Tevatron
Higgs search.

At the CERN LHC, however, gluon fusion will be the discovery channel
for the Higgs.  The cross section for light Higgs boson production
will be sufficiently large that the rare decay $H\to\gamma\gamma$ can
be used up to the point that the $WW^*$ channel begins to open up.
From that point on, the diboson decays provide a very robust signal.

\section{Methods}
The Higgs boson couples to mass, which presents a problem for hadronic
production.  Gluons are massless and therefore do not couple directly
to the Higgs at all, while the quarks that make up the proton have
very tiny masses.  Therefore, the dominant production mechanism is
gluon fusion via virtual top quark loops.  In the limit that the top
quark is very heavy, we can integrate out the top and formulate an
effective Lagrangian coupling the Higgs boson to the light quarks and
gluons~\cite{Shifman:1979eb,Vainshtein:1980ea,Voloshin:1986tc}.  If we
take the light quarks to be massless, the effective Lagrangian takes
the form
\begin{equation}
  {\cal L}_{\rm eff} = C_1(\alpha_s) H G^{a\mu\nu}G^a_{\mu\nu},
  \label{P5_kilgore_0706_eq1}
\end{equation}
where $G^a_{\mu\nu}$ is the gluon field strength tensor.  The
coefficient function $C_1(\alpha_s)$ has been computed to order
$\alpha_s^4$\cite{Chetyrkin:1997iv}.

The use of the effective Lagrangian allows us to replace massive loop
diagrams with point-like interactions.  Next-to-leading order (NLO)
corrections to inclusive Higgs production have been computed using
both the effective Lagrangian~\cite{Dawson:1991zj,Djouadi:1991tk} and
the full theory~\cite{Graudenz:1993pv,Spira:1995rr}.  One expects that
the effective Lagrangian will work very well if the Higgs mass is much
smaller than twice the top mass but that it will be unreliable for
larger masses.  In fact, it was found that at NLO the effective
Lagrangian does indeed agree very well with the full calculation below
the top threshold and was even found to agree to within $10\%$ for
Higgs masses as large as $1$ TeV.

It was also found that the NLO corrections are very large, of order
$70-100\%$.  Such large corrections clearly call for the evaluation of
still higher-order terms in order to arrive at a solid theoretical
understanding of the process.  Since the effective Lagrangian seems to
be a valid approximation, especially in the phenomenologically
interesting region of Higgs boson masses below $200$ GeV, we have
embarked on an effort to compute the full next-to-next-to-leading
order (NNLO) corrections in the heavy top limit.  In this talk, we
will present results for soft plus virtual corrections to inclusive
Higgs
production~\cite{Harlander:2000mg,Harlander:2001is,Catani:2001ic}.
These terms are not expected to dominate the full result and for this
reason we also discuss an approximation of the formally sub-leading
but numerically dominant contribution~\cite{Kramer:1998iq}.

\section{The Soft Approximation and Beyond}

There are three distinct contributions to inclusive Higgs production
at NNLO (see Figure~\ref{P5_kilgore_0706_fig1}): Virtual corrections to two
loops, single real radiation to one loop and double real radiation at
tree level.  \newdimen\figwid \figwid=\columnwidth \divide\figwid by4
\begin{figure*}
\includegraphics[width=\figwid]{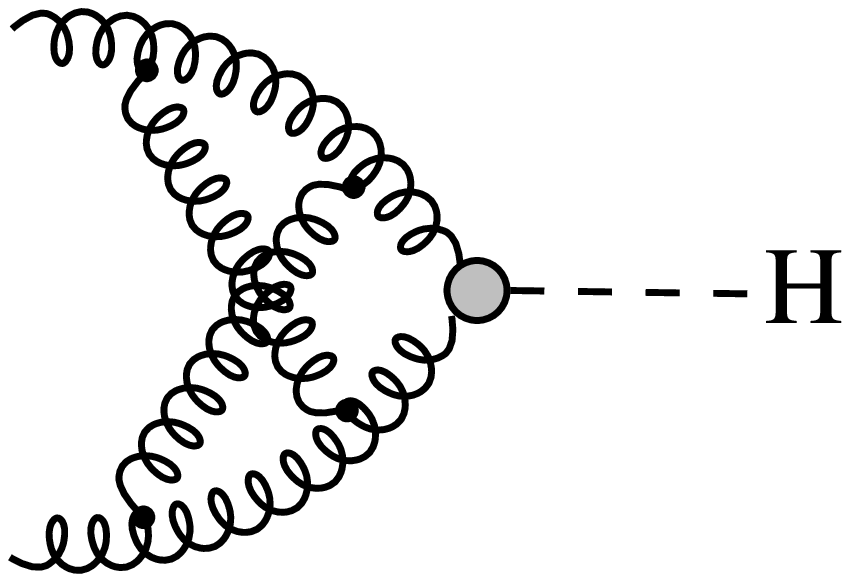}%
\hfil
\includegraphics[width=\figwid]{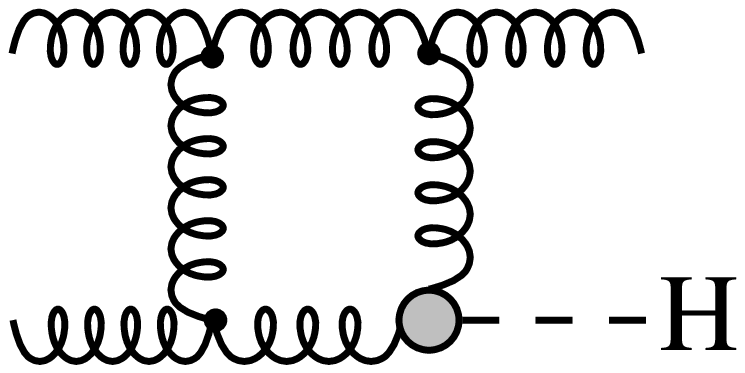}%
\hfil
\includegraphics[width=\figwid]{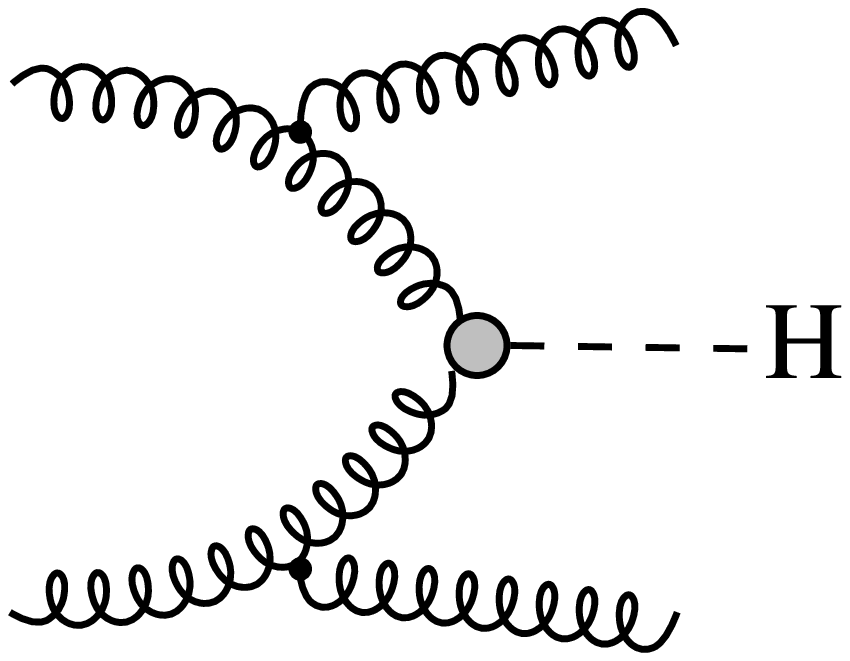}%
\caption{Representative diagrams of virtual corrections, single real
radiation and double real radiation.}
\label{P5_kilgore_0706_fig1}
\end{figure*}
These three channels produce radiative corrections that fall into
three categories, depending on their functional dependence on the
fraction $x\equiv M_H^2/\hat{s}$ of the center-of-mass energy squared
of the scattering process that goes into producing the Higgs boson.
\begin{equation}
  \sigma^{NNLO} =\ x\left[a\,\delta(1-x) + \sum_{n=0}^3
          b_n\left(\frac{\ln^n(1-x)} {1-x}\right)_+ + \sum_{n=0}^3 c_n
          \ln^n(1-x) + \dots\right].
  \label{P5_kilgore_0706_eq2}
\end{equation}
In the virtual corrections, all of the energy goes into Higgs boson
production, so these terms contribute only to the $\delta(1-x)$
correction.  Real emission processes generate terms like
$(1-x)^{n-m\epsilon}$ where $n\ge -1$ is an integer and $\epsilon$ is
the dimensional regularization parameter where space-time is taken to
be $D=4-2\epsilon$ dimensions.  These processes contribute to the $a$
and $b_n$ coefficients in equation~(\ref{P5_kilgore_0706_eq2}) by
expanding terms like $(1-x)^{-1-m\epsilon}$ as distributions
\begin{equation}
  (1-x)^{-1-m\epsilon} =\ -\frac{\delta(1-x)}{m\epsilon} +
      \sum_{n=0}^\infty \frac{(-m\epsilon)^n}{n!}\left[
      \frac{\ln^n(1-x)}{1-x}\right]_+.
  \label{P5_kilgore_0706_eq3}
\end{equation}

In the soft limit, there would be no energy carried away by real
emission and only the $\delta(1-x)$ term would be kept.  However, the
$\left[\frac{\ln^n(1-x)}{1-x}\right]_+$ terms are directly connected
to the $\delta(1-x)$ terms through
equation~(\ref{P5_kilgore_0706_eq3}) and in canceling the infrared
poles proportional to $\delta(1-x)$ we get these terms for free so
they are kept as part of the soft approximation.

While the soft approximation keeps the formally leading terms, it was
found that at NLO it is a poor approximation.  It is actually the
sub-leading $c_n$ ($n=0,1$ at NLO) terms that dominate the cross
section.  At NNLO, the $c_n$ terms are again expected to dominate.
Kr\"amer, Laenen and Spira~\cite{Kramer:1998iq} have used collinear
resummation to derive {\it approximate} NNLO results for $a, b_n$ and
$c_n$.  We expect their resummation to give the correct values for the
coefficients $b_3$, $b_2$ and $c_3$.  The other coefficients require
additional calculation, higher order resummation coefficients or, for
the remaining $c_n$, receive non-collinear contributions and we do not
expect the approximation to be accurate.  For the $a$ and $b_n$ terms
which we have computed directly, these expectations are fulfilled,
giving us confidence that the dominant term, $c_3$ is indeed accurate.

This gives us a range of possibilities for estimating the full NNLO
correction.  In Figure~\ref{P5_kilgore_0706_fig2} we show three
approximations in addition to the soft limit:\\
 1) Use $c_3$, $c_2$, $c_1$ and $c_0$ from Ref.~\cite{Kramer:1998iq},\\
 2) Use $c_3$ from Ref.~\cite{Kramer:1998iq} and generate sub-leading
    $\ln^n(1-x)$ terms by expanding $x b_n\to b_n + (1-x)b_n$,\\
 3) Use $c_3$ from Ref.~\cite{Kramer:1998iq} and drop all sub-leading
    $\ln^n(1-x)$ terms.\\
Note that in order to truly estimate the NNLO cross section, one needs
NNLO parton distribution functions (PDFs).  Unfortunately, the
necessary ingredients for producing NNLO PDFs are still being
developed.  Approximate NNLO PDFs have been produced, but at the time
of this work they are not yet publicly available.  We therefore use
the CTEQ5 NLO parton distribution functions~\cite{Lai:1999wy} and
acknowledge the inconsistency.

There are two outstanding features of
Figure~\ref{P5_kilgore_0706_fig2}: the formally sub-leading
$\ln^3(1-x)$ terms dominate the corrections, and the size of the
corrections is very large.  One expects that using NNLO PDFs will
reduce the magnitude of the correction by $\sim10\%$, but it will
still be very large.  We can take the spread between these
approximations as an estimate of the uncertainty due to the
uncalculated terms.

\section{Conclusions}
We have described a calculation of the soft plus virtual NNLO
corrections to inclusive Higgs production and estimates of the full
NNLO correction based on collinear resummation.  While the soft plus
virtual terms are perturbatively well-behaved, the leading non-soft
terms dominate the cross section and give rise to very large
corrections.  At this time, the two most important questions
concerning inclusive Higgs production are 1) What is the precise value
of the NNLO K-factor? and 2) How reliable is the NNLO K-factor with
respect to even higher order corrections?  The first question can be
answered by completing the full NNLO calculation and this work is
underway~\cite{RVHWBK:IP}.  The second question, which is crucial for
determining the precision to which the properties of the Higgs boson
can be measured at the LHC, requires further investigation.

\newdimen\figwid
\figwid=\columnwidth
\multiply\figwid by 4
\divide\figwid by 9
\begin{figure*}
\includegraphics[width=\figwid]{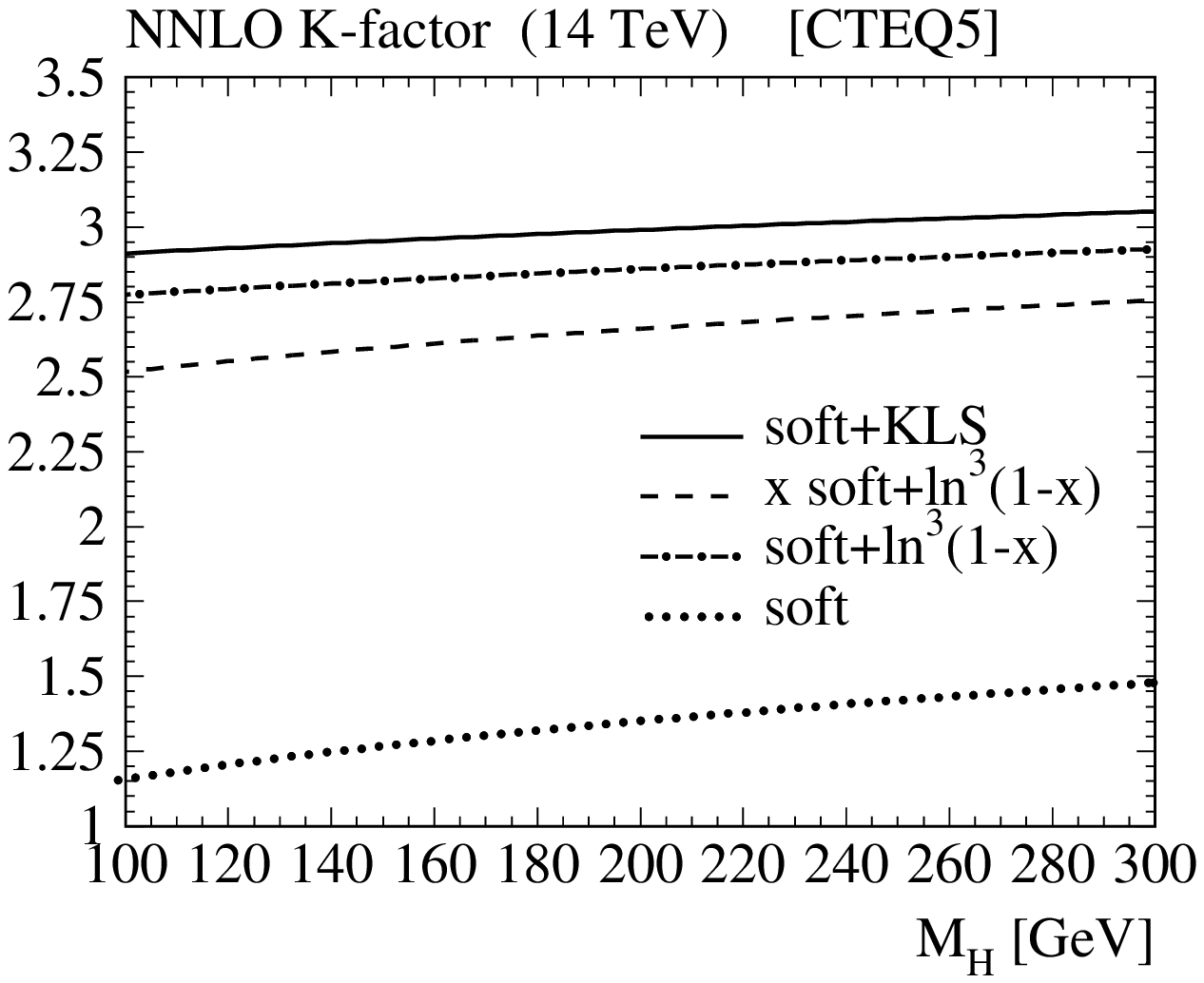}%
\hfil
\includegraphics[width=\figwid]{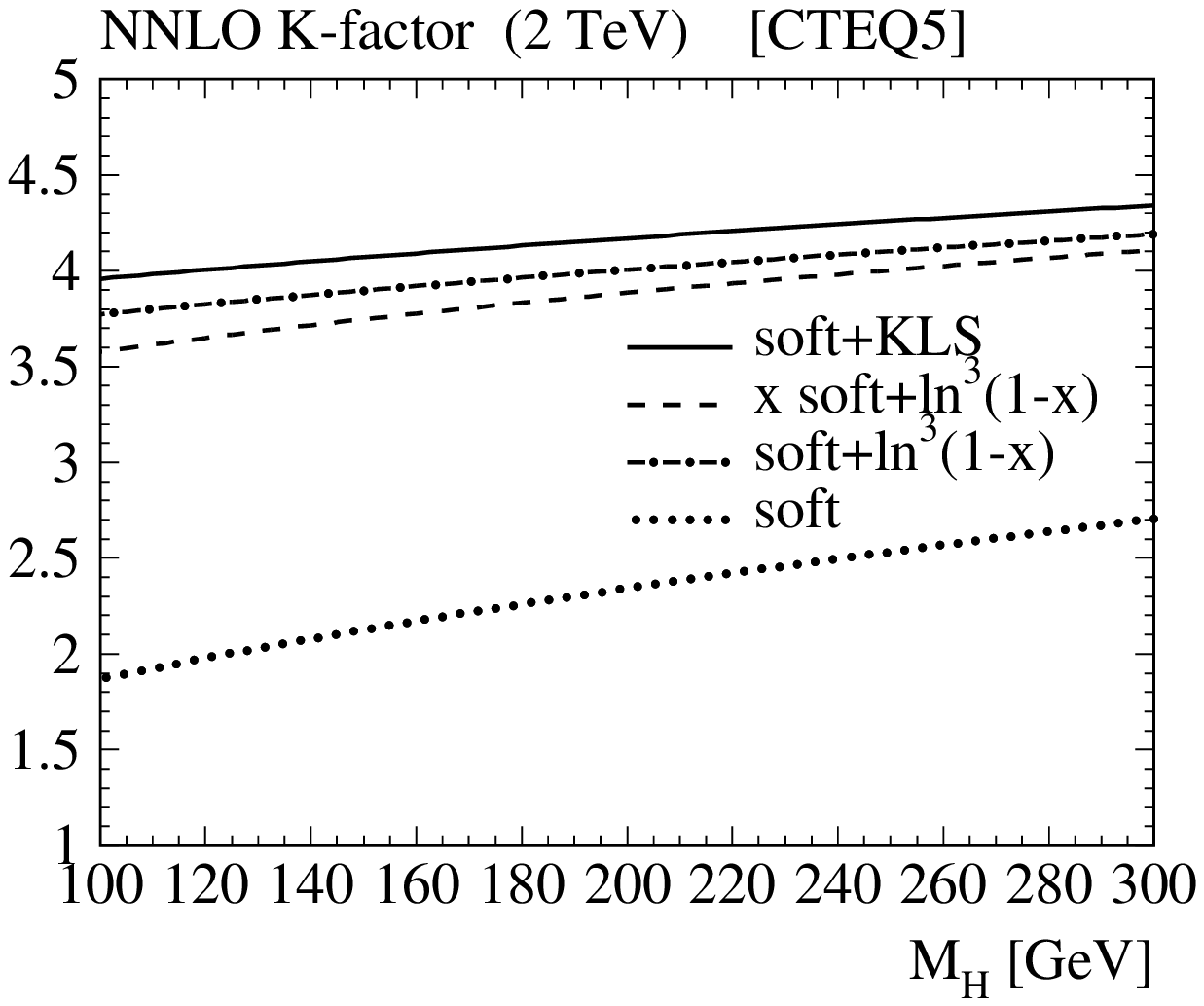}%
\caption{NNLO K-factor as a function of Higgs boson mass at (a)
$\sqrt{\hat{s}}=14$ TeV and (b) $\sqrt{\hat{s}}=2$ TeV.  The solid
line corresponds to approximation 1 above, the dashed line to
approximation 2 and the dash-dot line to approximation 3.  The dotted
line represents the soft approximation.}
\label{P5_kilgore_0706_fig2}
\end{figure*}


\end{document}